\documentclass[11pt,twoside]{article}

\usepackage{asp2006}
\usepackage{epsf}

\begin{document}
\title{The Star Formation Histories of the M31 and M33 Spheroids}   
\author{Thomas M. Brown}   
\affil{Space Telescope Science Institute, 3700 San Martin Drive, Baltimore,
MD 21218 USA}    

\begin{abstract} 
I review the observational constraints on the star formation histories
in the spheroids of M33 and M31, the other two spiral galaxies in the
Local Group.  M33 does not possess a traditional bulge; instead, it
has a small nuclear region hosting stars with a wide range of ages.
The star formation history of the M33 halo is poorly constrained, but
composite spectra of its halo globular clusters imply a wide age
spread of 5--7 years, while the presence of RR Lyrae stars in the halo
implies at least some of the population is ancient.  Although it is
possible to obtain the detailed star formation history of the M33 halo
via deep photometry, this has not been done to date.  M31 hosts a
traditional bulge that is apparently dominated by stars older than
10~Gyr.  Deep photometry of the M31 halo demonstrates that it hosts
both a population of ancient metal-poor stars and a significant
population extending to younger ages and high metallicity, apparently
due to its active merger history.
\end{abstract}


\section{Introduction}   

The Local Group hosts three spiral galaxies: the Milky Way, M31, and
M33.  The Milky Way and M31 are the only massive galaxies in the Local
Group, and there are indications that M31 is more representative of
massive spirals than the Milky Way \citep[e.g.,][]{fh07}.  While M33
is the third most massive galaxy in the Local Group, it is a distant
third \citep{svdb00}, and it is representative of the most common type
of spiral galaxy in the local universe \citep[see][]{cm99}.  Because
M31 and M33 are at respective distances of 770 kpc \citep{fm90} and
860 kpc \citep{as06}, we can obtain photometry of their resolved stellar
populations and thus constrain their star formation histories at a
fidelity exceeding that possible in the Milky Way, where such studies
of the field population are often hampered by distance and reddening
uncertainties.  Here, I review the observational constraints on the
spheroid (bulge and halo) populations of these galaxies.

\section{Age Constraints}   

Spheroids are generally dominated by stars older than 1~Gyr.  In such
populations, changes with age are less dramatic than they are for
younger systems.  The best age diagnostics come from photometry
reaching low-mass ($\sim$0.8 $M_\odot$) main sequence (MS) stars; such
photometry enables the reconstruction of the complete star formation
history with an age resolution of $\sim$1~Gyr, but it is difficult to
obtain such photometry outside the Milky Way system, due to crowding
and depth limitations.  Age constraints are also available through
photometry of later evolutionary phases, such as the horizontal branch
(HB), asymptotic giant branch (AGB), and red giant branch (RGB);
relative to MS stars, these brighter stars can be detected
in more distant and crowded regions, but the age resolution is poorer,
allowing one to distinguish between young ($<$3~Gyr),
intermediate-age (3--8 Gyr), and old ($>$8--13~Gyr) stars.

\section{M33 Spheroid}

The existence of a spheroid in M33 has been controversial for decades.
While the galaxy does not appear to have a bulge, it definitely
possesses a halo.  I discuss these components in turn below.

\subsection{M33 Bulge}

M33 apparently does not possess a bulge in the classical sense of that
term \citep{gdb92}, although this has been the subject of debate
\citep[e.g.,][]{mor93}.  Semantics aside, the galaxy hosts a small
nucleus that can be fit with a bulge profile that dominates within
$\sim$0.1$^\prime$ of the galaxy center \citep{sf02}.  In a $K$
vs. $V-K$ color-magnitude diagram (CMD) of the brightest stars in the
nucleus, \citet{sf02} find young, intermediate-age, and old stars with
a mean metallicity [Fe/H]=$-0.26$.  It is not possible to obtain photometry of
the low-mass MS stars in the nucleus with any observatory in operation
or in development, so the constraints on the star formation history will
be poor for the foreseeable future.

\subsection{M33 Halo}

The existence of a halo in M33 was also controversial in the past,
such that the galaxy was sometimes referred to as a ``pure disk''
galaxy.  However, in recent years a preponderance of evidence
demonstrated beyond any doubt that M33 hosts a stellar halo.  Although
some regions of the disk have {\it Hubble Space Telescope (HST)}
photometry reaching the stellar MS (Program 10190; PI Garnett), the
halo-dominated regions beyond the disk have not been imaged at
sufficient depth, so the star formation history of the M33 halo
remains poorly constrained.

\citet{rc02} kinematically segregated the stellar clusters of M33 into
disk and halo components.  Unresolved photometry and spectroscopy of
the clusters enabled a rough age estimate that is prone to
degeneracies between blue MS stars and blue HB stars.  The young
clusters in the sample have motions consistent with disk membership,
but the old clusters show a much larger velocity dispersion,
implying that 85\% belong to the halo and 15\% belong to the disk
(Figure 1).  The clusters with halo kinematics have a spread in
relative ages of 5--7~Gyr, much greater  than the globular cluster system
in the Milky Way halo.  Subsequent photometry of one globular
cluster in the sample was sufficiently deep to demonstrate that its
blue spectrum is due to the presence of blue MS stars and not blue HB
stars, proving its age is 5--8~Gyr than the other globular
clusters in their sample, but unfortunately the cluster cannot be
kinematically assigned to either the disk or halo \citep{rc06}.

\begin{figure}[!ht]
\plotfiddle{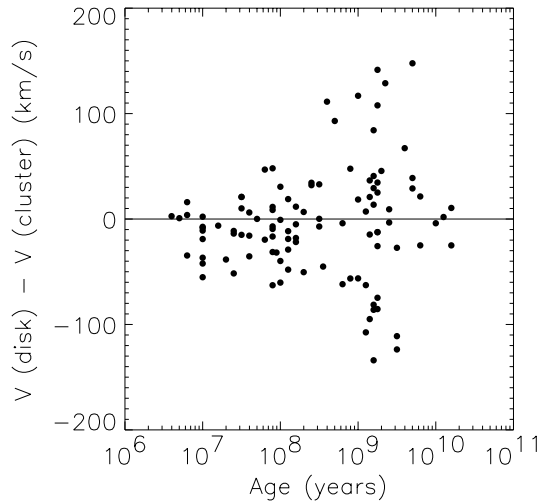}{2.4in}{0}{80}{80}{-130}{-410}
\caption{Difference between local disk velocity and measured cluster
velocity as a function age in M33 clusters \citep{rc02}, where cluster age 
is estimated from integrated photometry and spectroscopy.  The spread
in age for globular clusters in the M33 halo is much larger than it is
for such clusters in the Galactic halo. Plot provided courtesy of
R.\ Chandar.}
\end{figure}

\citet{as06} found evidence for a halo using 64 RR Lyrae variables
they identified in the galaxy.  They estimated reddenings from their
minimum $V-I$ colors and metallicities from their periods.  The
resulting distributions of reddening and metallicity imply they belong
to two distinct populations, associated with the disk and halo,
providing evidence for these components in the field that complements
the evidence in clusters (above).  RR Lyrae stars are only present in
ancient ($>$10~Gyr) populations, implying that both the M33 disk and
halo host at least some ancient stars.  In the [Fe/H] distribution for
the halo RR Lyrae stars, the peak is at $-1.3$, although the
metallicities of the RR Lyrae stars are not necessarily representative
of the underlying population, because metallicity affects the HB
temperature distribution and thus the fraction of HB stars falling in
the instability strip.

\citet{awm06} also found evidence for a halo field population.  From
spectroscopy of 280 stars, they were able to segregate their sample
into three components using both kinematics and metallicity: the halo,
the disk, and a third unknown component possibly associated with a stellar
stream.  The halo component exhibits a mean [Fe/H] of $-1.5$ and a
velocity dispersion of $\sim$50 km s$^{-1}$, but the data provide no
age constraints.

Using star counts of RGB and AGB stars along the minor axis,
\citet{mt08} found a break in the surface brightness profile at
11~kpc, where the profile changes from that of an exponential disk to
a power-law halo \citep[see also][]{mt09}.  The CMD of these bright
stars implies the halo population is dominated by stars older than
3~Gyr with a mean [Fe/H] of $-1.2$.  This is in good agreement with
the CMD of \citet{bwh04}, who obtained $V$ and $I$ photometry in the
halo outskirts and found a mean [Fe/H] of $-1.24$.

The next logical step in the investigation of the M33 halo is deep
photometry reaching its low-mass MS stars.  This is the only way to
unambiguously characterize the age distribution in the field
population.  While such photometry has been obtained in multiple
regions of the M31 halo with {\it HST} (discussed below),  the M33
halo remains unexplored, with very loose constraints on its star formation
history. 

\section{M31 Spheroid}

While M31 certainly has both a bulge and a halo, the distinction
between these components is somewhat muddled in the literature.
Historically, studies of the bulge and halo were driven by the
appearance of M31 in wide-field shallow imaging \citep[e.g.,][]{wk88},
with ``bulge'' studies generally focusing on the region within a kpc
of the center, and ``halo'' studies focusing on regions beyond 10~kpc
on the minor axis.  However, subsequent studies showed that the
spheroid looks like a bulge out to $\sim$20~kpc in both its
surface-brightness profile \citep{pvdb94} and metallicity distribution
\citep{mk86,dhp94,dhp01}.  It was only recently discovered that the
surface-brightness profile transitions to a power-law
\citep{pg05,mji05} and the metallicity drops by $\sim$1 dex
\citep{jsk06} beyond 20~kpc, as expected for a halo.  For consistency
with historical studies, I will refer to all regions beyond 10~kpc on
the minor axis as ``halo'' despite the persistence of some bulge-like
properties out to 20~kpc.

\subsection{M31 Bulge}

Two fields in the M31 bulge were recently imaged in the near-IR at
high resolution using adaptive optics on Gemini North \citep{tjd05},
resolving the upper 4--5~mag of the RGB and AGB.  The resulting $H$
and $K$ CMDs are consistent with a population dominated by stars at an
age of $\sim$10~Gyr, with a metallicity near solar, although the best-fit
models include a minority intermediate-age component that may be
spurious \citep{kago06}.  As with the center of M33, no observatory
in operation or development is capable of obtaining the detailed 
star formation history of the M31 bulge, because crowding precludes
photometry of the low-mass MS stars.

\subsection{M31 Halo}

Until the past decade, studies of the resolved stellar populations in
the M31 halo generally focused on the metallicity distribution
\citep[e.g.,][]{mk86,dhp94,dhp01}.  Metallicity distributions were
usually fit by assuming an old age ($>$10~Gyr) and then comparing the
stars on the AGB and RGB to isochrones or globular cluster templates,
although some studies were deep enough to reach the HB
\citep[e.g.,][]{hfr96}.  Although the RGB, AGB, and HB distributions
are, in principle, sensitive to age in broad age bins (as noted
above), several factors prevented these studies from exploring the age
distribution, including photometric scatter, insufficient star counts,
and contamination from foreground Milky Way dwarfs.  

When the Advanced Camera for Surveys (ACS) was installed on {\it HST}
in 2002, it became possible to obtain photometry of low-mass MS stars
in the M31 halo, enabling the exploration of both the metallicity and
age distributions.  \citet{tmb03} imaged a field 11~kpc from the
nucleus on the southeast minor axis, obtaining photometry of 250,000
stars down to $V\sim30.5$~mag in bands similar to $V$ and $I$ (Figure 2).  
By providing a large number of stars with small photometric
errors on the low-mass MS, the catalog was immune to contamination
from foreground Milky Way dwarfs.  The resulting fit to the CMD found
a wide age range in addition to the wide metallicity range that was
already known, and speculated that this was due to a significant
merger or series of smaller mergers in the galaxy.  In the best-fit
model \citep{tmb06}, $\sim$40\% of the stars are younger than 10~Gyr
and more metal-rich than 47~Tuc ([Fe/H]~=~--0.7), with significant
numbers of stars down to ages of 2~Gyr.  Besides the field population,
there is evidence from integrated colors and spectroscopy that the M31
globular cluster system also extends to intermediate ages
\citep{thp05,mab05}.

\begin{figure}[ht!]
\plotone{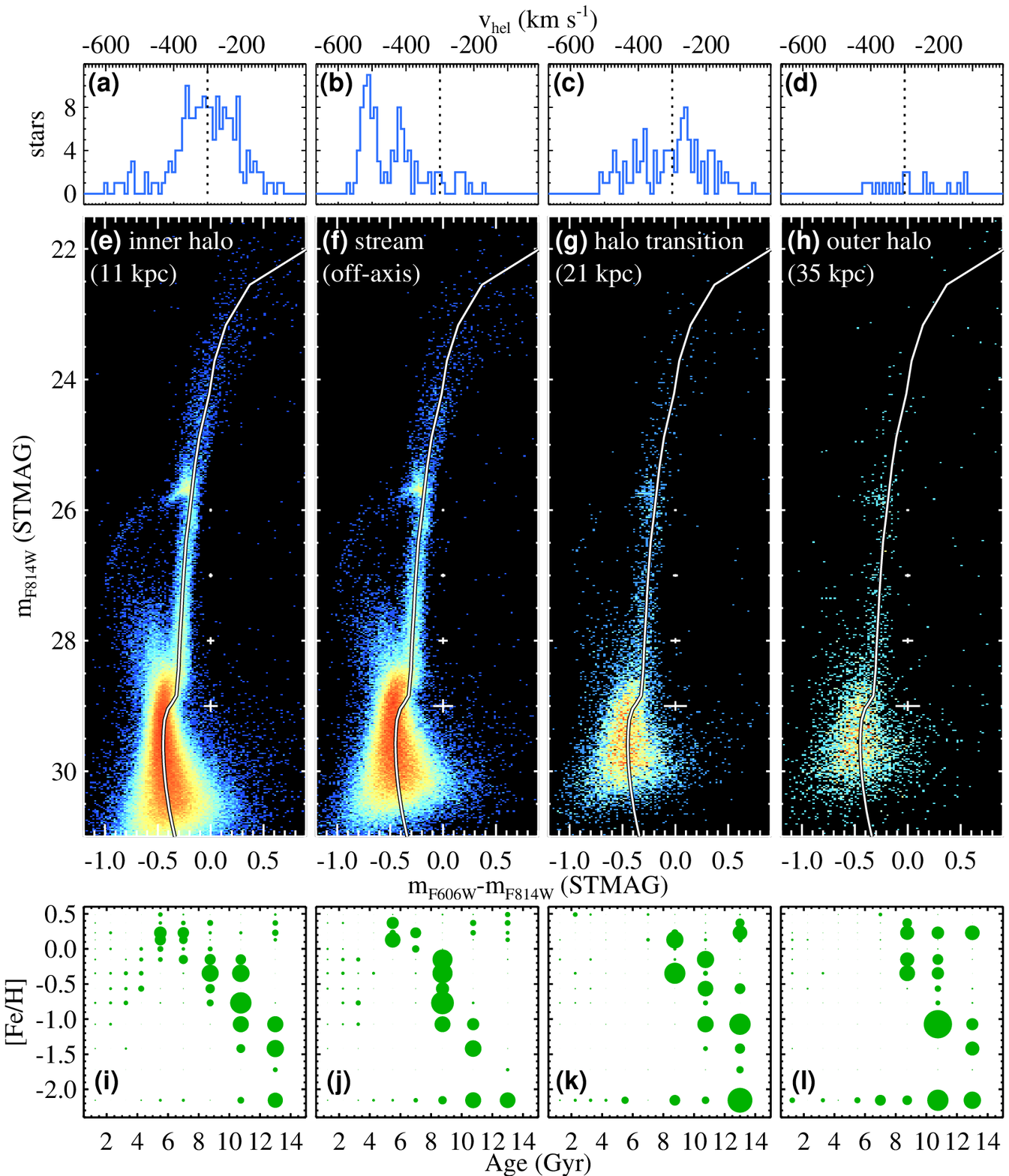}
\caption{Radial velocities (top row; dotted line is M31 systemic
  velocity), CMDs (middle row; curve shows 47 Tuc ridge line for
  comparison), and star formation histories (bottom row; area of
  circles proportional to weight in fit) for four regions in the M31
  halo \citep{tmb03,tmb06,tmb07,tmb08}.  The three regions on the
  minor axis (at 11, 21, and 35~kpc) show a kinematically hot
  population at the M31 systemic velocity, while the off-axis field
  shows the kinematically cold stream plunging into M31 toward us from
  behind the galaxy.  Despite their distinct kinematic profiles, the
  CMDs and associated star formation histories for the 11 kpc field
  and the stream are very similar, due to the inner halo being
  polluted by debris from the stream's progenitor
  \citep{tmb06,maf07,kmg07}.  Although the fields further out on the
  minor axis do not include significant numbers of stars as young and
  metal-rich as those found in the 11~kpc and stream fields, all of
  the halo fields exhibit an extended star formation history,
  consistent with expectations from hierarchical merging.}
\end{figure}

The recent wide-field imaging surveys of M31 clearly show that it has
undergone a violent merger history \citep{amnf02,ri07}, including a
giant stellar stream resulting from the tidal debris of a recent
merger event \citep{ri01}.  Subsequent studies demonstrated that the
inner spheroid of M31 (within $\sim$15~kpc) is polluted by material
stripped from progenitor satellite of the giant stellar stream.  This
evidence includes N-body simulations of the satellite disruption that
reproduce the morphology of the major substructures in the galaxy
\citep{maf07}, kinematical surveys that confirm the motions in the
N-body simulations \citep{kmg07}, and followup ACS imaging of the
giant stellar stream that shows strong similarities between the star
formation histories of the stream and inner 
spheroid \citep[Figure 2;][]{tmb06}.

Given the discovery that the M31 halo becomes more like a halo beyond
20~kpc, a subsequent ACS survey explored regions on the minor axis
further out, at 21~kpc \citep[in the transition region;][]{tmb07} and
35~kpc \citep[where the spheroid clearly exhibits a halo surface
  brightness profile and metallicity;][]{tmb08}.  Compared to the
field at 11~kpc, these fields host far fewer stars at ages younger
than 8~Gyr, but the populations clearly do not represent a classical
halo formed via monolithic collapse at early times (Figure 2); in the
best-fit model, $\sim$30\% of the stars are younger than 10~Gyr, and
only $\sim$10\% of the stars are ancient ($\ge$12 Gyr) and metal-poor
([Fe/H]$\le$--1.5).  All regions of the halo explored to date are
consistent with a history whereby the galaxy forms over a prolonged
period of hierarchical merging.

\section{Summary}

M33 possesses a halo but not a traditional bulge.  This halo exhibits
secondary evidence for intermediate-age populations, such as a halo
globular cluster system spanning a wide age range in integrated
spectroscopy \citep{rc02}, but the age distribution in the M33 halo
has not been constrained via photometry of the low-mass MS stars, as
done in M31.  Thus, our understanding of the star formation history in
the M33 halo is in a state similar to that found in M31 prior to the
advent of the {\it HST} ACS.  For years, M31 was assumed to host an
old halo of age $>$10~Gyr.  The M31 halo has now been probed in
multiple locations along the minor axis, spanning the regions where
the halo looks more like a bulge and where it looks more like a
traditional halo, but in all of these regions it exhibits an extended
star formation history \citep{tmb06,tmb07,tmb08}.  The M33 halo
remains the last spiral galaxy halo unexplored via photometry of its
low-mass MS stars, leaving its star formation history poorly
constrained.  Appropriately deep imaging of the M33 halo should be
obtained during the remaining {\it HST} mission, or it may be many
years before such data can be obtained again.


\acknowledgements 
I am grateful to my collaborators on the {\it HST}/ACS M31 observing
programs for their contributions to those projects, and to R.\ Chandar
for useful discussions and suggestions.


\end{document}